\documentclass{cupconf}
\usepackage{psfig}
  \checkfont{eurm10}
  \iffontfound
    \IfFileExists{upmath.sty}
      {\typeout{^^JFound AMS Euler Roman fonts on the system,
                   using the 'upmath' package.^^J}%
       \usepackage{upmath}}
      {\typeout{^^JFound AMS Euler Roman fonts on the system, but you
                   dont seem to have the}%
       \typeout{'upmath' package installed. cupconf.cls can take advantage
                 of these fonts,^^Jif you use 'upmath' package.^^J}%
      }
  \else
  \fi

  \checkfont{msam10}
  \iffontfound
    \IfFileExists{amssymb.sty}
      {\typeout{^^JFound AMS Symbol fonts on the system, using the
                'amssymb' package.^^J}%
       \usepackage{amssymb}%
         \let\leq=\leqslant
         \let\geq=\geqslant
      }{}
  \fi

  \IfFileExists{amsbsy.sty}
    {\typeout{^^JFound the 'amsbsy' package on the system, using it.^^J}%
     \usepackage{amsbsy}}
    {}

\newsavebox{\astrutbox}
\sbox{\astrutbox}{\rule[-5pt]{0pt}{20pt}}

\newcommand\etal{\mbox{\textit{et al.}}}

\newcommand\eg{e.g.\ }

\newcommand{\hst}{\textit{HST}}
\newcommand{\spitzer}{\textit{Spitzer}}

\newcommand{\iso}{\textit{ISO}}

\newcommand{\lsim}{\lesssim}
\newcommand{\gsim}{\gtrsim}

\newcommand{\lsol}{\hbox{$L_\odot$}}
\newcommand{\lstar}{\hbox{$L^\ast$}}

\newcommand{\apj}{ApJ}
\newcommand{\apjl}{ApJL}
\newcommand{\apjs}{ApJS}
\newcommand{\aj}{AJ}
\newcommand{\mnras}{MNRAS}
\newcommand{\aap}{A\&A} 
\newcommand{\araa}{ARAA} 
 
\newcommand{\micron}{\hbox{$\mu$m}}

\newcommand{\arcsec}{\mbox{$^{\prime\prime}$}}
\newcommand\phs{\phantom{-}}
\newcommand\phn{\phantom{0}}

\title[Distant Infrared--Luminous Galaxies]{Studying Distant Infrared--Luminous
  Galaxies with \textit{Spitzer} and \textit{Hubble}}

\author[C. Papovich \etal]{%
C\ls A\ls S\ls E\ls Y\ns P\ls A\ls P\ls O\ls V\ls I\ls C\ls H,\ns 
E\ls I\ls I\ls C\ls H\ls I\ns E\ls G\ls A\ls M\ls I,\ns \\
E\ls M\ls E\ls R\ls I\ls C\ns L\ls E\ns F\ls L\ls O\ls C'\ls H,\ns
P\ls A\ls B\ls L\ls O\ns P\ls \'E\ls R\ls E\ls Z\ls \hbox{-}\ls G\ls O\ls N\ls
Z\ls\'A\ls L\ls E\ls Z,\ns \\
G\ls E\ls O\ls R\ls G\ls E\ns R\ls I\ls E\ls K\ls E,\ns
J\ls A\ls N\ls E\ns R\ls I\ls G\ls B\ls Y,\ns
H\ls E\ls R\ls V\ls \'E\ns D\ls O\ls L\ls E,\ns \\ \and
M\ls A\ls R\ls C\ls I\ls A\ns R\ls I\ls E\ls K\ls E
}

\affiliation{Steward Observatory, University of Arizona, 933 N. Cherry
Avenue, Tucson, AZ 85741, USA}

\pubyear{2004}
\volume{1}
\pagerange{1--11}
\date{\today and in revised form \today}
\begin{document}

\maketitle

\begin{abstract}
New surveys with the \spitzer\ space telescope identify distant
star--forming and active galaxies by their strong emission at
far--infrared wavelengths, which provides strong constraints on these
galaxies' bolometric energy. Using early results from \spitzer\
surveys at 24~\micron, we argue that the faint sources correspond to
the existence of a population of infrared--luminous galaxies at
$z\gsim 1$ that are not expected from predictions based on previous
observations from \iso\ and \textit{IRAS}.   Combining \spitzer\
images with deep ground--based optical and \textit{Hubble} Space
Telescope imaging, we discuss the properties of galaxies selected at
24~\micron\ in the region of the \textit{Chandra} Deep Field South,
including redshift and morphological distributions.   Galaxies with
$z\lsim 1$ constitute roughly half of the faint 24~\micron\ sources.
Infrared--luminous galaxies at these redshifts span a wide variety of
normal to strongly interacting/merging morphologies, which suggests
that a range of mechanisms produce infrared activity.  Large--area,
joint surveys between \spitzer\ and \hst\ are needed to understand the
complex relation between galaxy morphology, structure, environment and
activity level, and how this evolves with cosmic time.   We briefly
discuss strategies for constructing surveys to maximize the legacy of
these missions.

\end{abstract}

\firstsection % if your document starts with a section,
\section{Introduction\label{sec:intro}}

Infrared (IR) luminous galaxies represent highly active stages in
galaxy evolution that are not generally inferred in optically selected
galaxy surveys (\eg Rieke \& Low 1972; Soifer, Neugebauer, \& Houck
1987). High IR--emission is typically generated in heavily enshrouded
starbursts associated with morphologically disturbed or merging
galaxies (Sanders \etal\ 1988); in comparison, optical studies probe
less obscured star--forming regions often located in galaxy disks (\eg
Kennicutt 1998).  At the present day, most of the light emitted from
galaxies comes at optical wavelengths, with only one--third of the
bolometric luminosity density coming in the IR (Soifer \& Neugebauer
1991).   However, the cosmic background implies that the far--IR
emission from galaxies in the early Universe is as important
energetically as the emission in the optical and UV combined (Hauser
\etal\ 1998), and IR number counts from \textit{ISO} indicate that the
these sources evolved faster than that inferred directly from
UV/optical observations.  The interpretation of these counts combined
with models of the cosmic IR background (Elbaz \etal\ 2002; Dole
\etal\ 2003) argues that IR--luminous stages of galaxy evolution were
frequently more common at high redshift.  As a result, IR--luminous
galaxies may be responsible for a substantial fraction of the
global star--formation and metal--production rate  (\eg Franceschini
\etal\ 2001).

Studying the mechanisms for this apparent rapidly evolving
IR--luminous galaxy population has been problematic, primarily due to
low--number statistics of sources at high redshifts, and difficulty in
measuring their multi--wavelength properties and internal structure.
The improvements in IR sensitivity and survey efficiency now possible
with {\it Spitzer} allow major advances in the study of the
IR--luminous stages of galaxy evolution, particularly in true
panchromatic datasets.  Measurements at 24~$\mu$m with the
\spitzer/MIPS instrument are key because of their sensitivity to IR
emission and angular resolution (see Rieke \etal\ 2004).  Deep,
large--area surveys with \spitzer\ are currently underway, and early
results demonstrate that \spitzer--selected IR--luminous galaxies are
not only  common at high--redshifts ($z \gsim 1$; \eg Le Floc'h \etal\
2004), but may bridge the gap between the far--IR/sub-mm galaxy
populations at $z\sim 3$ and galaxies at more moderate
redshifts --- placing them in the context of the cosmic
star--formation--rate density (\eg Egami \etal\ 2004).

In these proceedings, we present early results from 
IR surveys with \spitzer\ at 24~\micron\ from time allocated to the
Guaranteed Time Observers (GTOs),  and we consider implications
for IR--stages of galaxy evolution.  We then describe the properties of
24\micron--selected galaxies using deep ground--based and \hst\
imaging, and we discuss questions that arise when considering
IR--luminous galaxies within contemporary theories of galaxy evolution.
Wherever appropriate, we assume a cosmology with $H_0 = 70$~km s$^{-1}$
Mpc$^{-1}$, $\Omega_m=0.3$, and $\Lambda = 0.7$.

\section{\spitzer\ Observations of Distant IR--Luminous Galaxies}

%\subsection{\spitzer\ 24~\micron\ number counts}

\begin{figure*}
%XC\epsscale{1.111}
\centerline{\psfig {figure=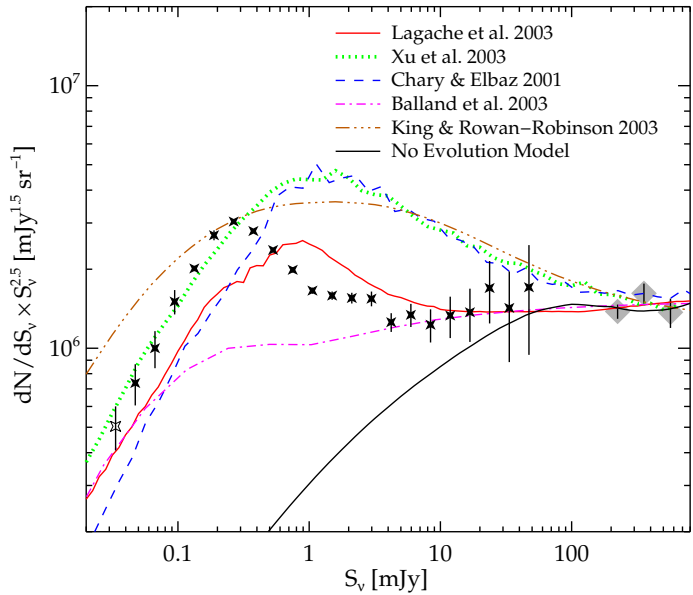,width=4in}}
\caption{ Differential \spitzer\ 24~\micron\ number counts (from
  Papovich \etal\ 2004),  normalized to a Euclidean slope, $dN/dS_\nu
  \sim S_\nu^{-2.5}$.  The solid stars show the average counts from
  all the \spitzer\ fields.  Each flux bin is $\Delta( \log S_\nu) =
  0.15$~dex.  The shaded diamonds correspond to \textit{IRAS}
  25~\micron\ number counts from Hacking \& Soifer (1991).  The curves
  show the predictions from various contemporary models from the
  literature (see figure inset; and adjusted slightly to match the
  observed \textit{IRAS} counts), and a model based on the local \iso\
  15~\micron\ luminosity function and assuming non-evolving galaxy
  SEDs.\label{fig:counts}}
\end{figure*}

\spitzer\ provides efficient, deep observations of large areas of sky
containing a high source surface density.  Figure~\ref{fig:counts}
shows the 24~\micron\ differential number counts that have been
derived using roughly 50,000 galaxies from five fields spanning
approximately 10.5 square degrees (Papovich \etal\ 2004).  At bright
flux densities, $S_\nu \gsim 5$~mJy, the differential 24~\micron\
source counts increase at approximately the Euclidean rate, $dN/dS_\nu
\sim S^{-2.5}$, which extends the trends observed by the IRAS
25~\micron\ population by two orders of magnitude (Hacking \& Soifer
1991; Shupe \etal\ 1998).  For $S_\nu \simeq 0.4-4$~mJy, the counts
increase at a super--Euclidean rate, and peak near $0.2-0.4$~mJy.
This observation is similar to the trend observed in the \textit{ISO}
15~\micron\ source counts (Elbaz \etal\ 1999), but the peak in the
24~\micron\ differential source counts occurs at fluxes fainter by a
factor $\approx 2.0$.  The counts converge rapidly at $\lsim 0.2$~mJy,
with a faint--end slope of $dN/dS_\nu \sim S_\nu^{-1.5\pm 0.1}$.

The thick line in figure~\ref{fig:counts} shows the expected counts
from non--evolving models of the local IR--luminous population.  While
the non--evolving fiducial model is consistent with the observed
24~\micron\ counts for $S_\nu \gsim 20$~mJy, it underpredicts the
counts at $S_\nu \lsim 0.4$~mJy by more than a factor of 10.  The
\spitzer\ 24~\micron\ number counts require strong evolution in the
IR--luminous galaxy population.  This is similar to conclusions based
on data from \textit{IRAS} and \textit{ISO}, but the \spitzer\ counts
extend them to fainter fluxes (and higher redshifts, see below) than
those probed from these earlier missions. 

\subsection{Interpretation of 24~\micron\ sources}

The 24~\micron\ source counts differ strongly from
predictions of various contemporary models (as labeled in
figure~\ref{fig:counts}).  Four of the models are phenomenological in
approach (so--called `backwards-evolution' models), which evolve the
parameters of the local luminosity function back in time (generally
accounting for density and luminosity evolution by changing
$\phi^\ast$ and \lstar) to match counts from \textit{ISO}, radio,
sub--mm, and other datasets. Several models (Chary \& Elbaz 2001; King
\& Rowan--Robinson 2001; Xu \etal\ 2003) show a rapid increase in the
number of sources with super--Euclidean rates at relatively bright
flux densities ($S_\nu \gsim 10$~mJy).  They predict 24~\micron\
counts that peak near 1~mJy, and overpredict the measured counts at
this flux density by factors of $2-3$.  These models expected there to
be  more luminous IR galaxies (LIRGs) and ultra--luminous IR galaxies
(ULIRGs) selected by \spitzer\ 24~\micron\ near $z\sim 1$, based
largely on the redshift distribution of the \textit{ISO}
15~\micron\ sources.  Lagache \etal\ (2003) predicted roughly
Euclidean counts for $S_\nu > 10$~mJy.  The shape of the counts in
this model is similar to the observed distribution, but it peaks at
$S_\nu \sim 1$~mJy, at higher flux densities than the observed counts.
This model has a redshift distribution that peaks near $z\sim 1$, but
tapers somewhat slower with a significant population 24~\micron\
sources out to $z\gsim 2$ (Dole \etal\ 2003). 

\textit{The peak in the 24~\micron\ differential number counts occurs at
fainter flux densities than predicted from the models based on the
\textit{ISO} results.}   Because the number counts are essentially just the
integral of the galaxy luminosity function over redshift and flux
down to the survey flux limit, they are likely dominated by galaxies
with `\lstar' luminosities (modulo variations in the faint--end slope
of the luminosity function).   Models that reproduce the IR
background require far--IR luminosity functions with
$\lstar(\mathrm{IR}) \gsim 10^{11}$~\lsol\ (see Hauser \& Dwek 2001).
Elbaz \etal\ (2002) observed that the redshift distribution of objects
with these luminosities in deep \textit{ISO} 15~\micron\ surveys spans
$z\simeq 0.8-1.2$, and that these objects constitute a large fraction
of the total cosmic IR background.  Assuming the 24~\micron\ number
counts at $0.1-0.4$~mJy correspond to \lstar\ galaxies, their
redshifts must lie at $z > 1$.

\subsection{Challenges to Galaxy Evolution Theories}

Recently, Lagache \etal\ (2004) have updated their phenomenological
model in order to reproduce the measured \spitzer\ number counts.  To
do this, they required a minor modification of the redshift
distribution of 24~\micron\ sources, such that galaxies with $z\gsim
1$ contribute more than half of the counts at faint fluxes ($\sim
0.2$~mJy).  They also required an adjustment to the flux density in
the mid--IR region of galaxy SEDs ($3-30$~\micron) of up to a factor
of two.  The implications are: 1) that stochastically heated emission
features at mid--IR wavelengths (UIBs and PAHs) likely persist at high
redshifts ($z\gsim 2$); and 2) the relative strength of the various
mid--IR features may evolve with redshift.  The second implication is
not wholly unexpected as higher redshift galaxies may have very
different metallicity and chemistry, and the cosmic UV radiation field
is more intense (the latter contributes to the heating of the grains
responsible for the mid--IR emission features, \eg\ D\'esert \etal\
1990).  The intriguing prospect is that the mid--IR SEDs of
IR--luminous galaxies may evolve with redshift, which complicates modeling
efforts.  Forthcoming spectroscopy of high--redshift galaxies at
mid--IR wavelengths with the \spitzer\ Infrared Spectrograph will
measure the strength of these features and will help to constrain any
evolution observationally.

Although backwards--evolution models provide a useful framework for
parameterizing the strong evolution of IR--luminous galaxies, they are
unable to explain the physics responsible for this evolution.  Models
of galaxy formation and evolution that start from first principles (so
called `forward--evolution' models) currently lack the means of
producing either the strong evolution observed in the IR number counts
or in the cosmic IR background (see, \eg Hauser \& Dwek 2001).  For
example, the dot--dashed line in figure~\ref{fig:counts} shows the
model of Balland, Devriendt, \& Silk (2003), which is based on
semi--analytical hierarchical models within the Press--Schecter
formalism.  In that model, galaxies identified as `interacting' are
assigned IR--luminous galaxy SEDs.   This model includes additional
physics in that the evolution of galaxies depends on their local
environment and merger/interaction histories.  Although this model
predicts a near--Euclidean increase in the counts for $S_\nu \gsim
10$~mJy, the counts shift to sub--Euclidean rates at relatively bright
flux densities.  Semi--analytic models by R.\ Somerville, J.\ Primack,
\etal\ (in preparation), which broadly reproduce optical--near-IR
properties of galaxies from $z\sim 0-3$, predict an IR background
intensity that is too faint by several factors.  These examples are
typical of the general status of forward--evolution modeling efforts.
\textit{We are faced with a lack of understanding why such rapid evolution
occurs in the IR--luminous galaxy population.}

\section{Ground--based Observations of Distant \spitzer\ Galaxies}

The \spitzer\ GTO extragalactic survey fields were selected to have low
zodiacal and Galactic backgrounds (see \S~5; table~1), and to have the
highest--quality ancillary data available at other wavelengths.  The
GTOs used \spitzer\ to observe a $1\times 0.5$ sq.\ degree region of
\textit{Chandra} Deep Field South (CDF--S)  in early February 2004.
The CDF--S has exceptional ancillary data from X--ray to radio
wavelengths.  For the remainder of this contribution we will discuss
only a portion of these data --- focusing on the optical imaging and
redshift distribution of the \spitzer--selected galaxies.  Studies of
the \spitzer\ sources in this field using other ancillary data have
been carried out or are in progress.  For example, Rigby
\etal\ (2004) study the properties of X-ray--selected \spitzer\ 24~\micron\
sources in this field.

The region around the CDF--S has been the target of  several
ground--based imaging surveys.  Of these, the COMBO--17 survey
(Classifying Objects by Medium--Band Observations in 17 filters; Wolf
\etal\ 2003) has observed a $30\times 30$ sq.\ arcmin region around
the CDF--S field with imaging from $0.3-1$~\micron.   Using a suite of
medium--band filters, they provide highly--reliable photometric
redshifts for galaxies with $R\leq 24$ to $z\lsim 1.3$, and for AGN
out to substantially higher redshift (Wolf \etal\ 2004).   Nearly the
entire COMBO--17 field overlaps with the \spitzer\ field.  Most
\spitzer\ 24~\micron\ sources brighter than 60~$\mu$Jy (the estimated
50\% completeness limit) are readily identified in the COMBO-17
images: 3850 of the 4720 24~\micron\ sources in this region have
optical counterparts to $R \leq 25$ within 2\arcsec\ (the \spitzer\
24~\micron\ PSF is roughly 6\arcsec\ FWHM).  Of these, roughly 2970
have good photometric redshift estimates. Several large spectroscopic
surveys from the VLT (Le Ferv\'e \etal\ 2004; Vanzetta \etal\ 2004)
provide  an additional 290 \spitzer\ 24~\micron\ sources.

\begin{figure}
\centerline{\psfig {figure=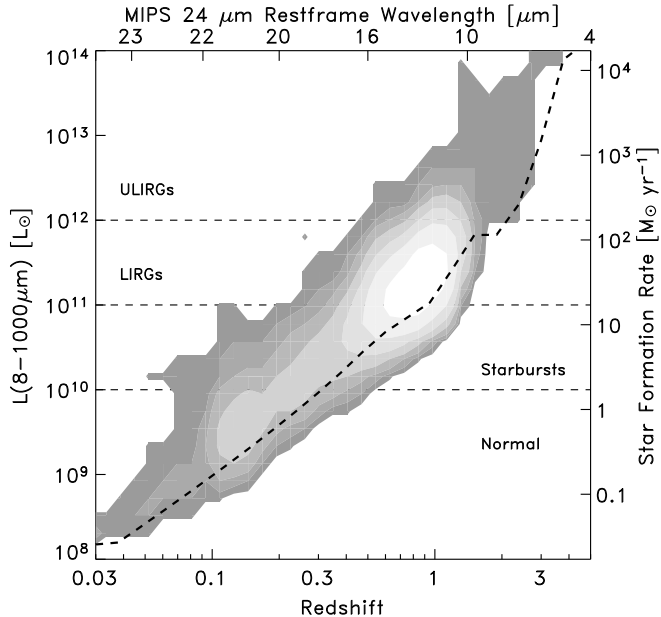,width=3.7in}}
\caption{Redshift and luminosity distribution of optically--selected
\spitzer/MIPS   24~\micron\ sources.  Redshifts correspond to values
published for   the COMBO-17 survey (Wolf \etal\ 2004), with
additional   spectroscopic redshifts from VLT/FORS2 (Vanzella \etal\
2004) and   VIRMOS (Le Ferv\'e \etal\ 2004).  Contours indicate
regions containing 1, 2, 4, 8, 16, 32, and 64 galaxies in bins of
$\Delta\log L(\mathrm{IR}) = 0.2$ and $\Delta\log z = 0.06$.   The heavy,
dashed line shows the estimated 80\% completeness limit of the
24~\micron\ imaging (see Papovich \etal\ 2004).  The right--hand axis
shows the SFR corresponding to $L(8-1000\micron)$ for the assumption
that all the IR luminosity results from star formation, and using the
relation established by Kennicutt (1998).  The top axis shows the
rest--frame wavelength observed at 24~\micron. Regions separated by
dashed lines show fiducial IR--galaxy classes.\label{fig:lumir}}
\end{figure}

\subsection{The redshift distribution of \spitzer\ 
24~\micron\ sources}   

Figure~\ref{fig:lumir} shows the redshift and
luminosity distribution of \spitzer--selected sources with
counterparts in the photometric-- and spectroscopic--redshift catalogs
from the CDF--S.   The total IR luminosity, $L(8-1000\micron)$, is
calculated by converting the measured 24~\micron\ flux density to a
luminosity using the reported redshift, then extrapolating to the
total IR luminosity using the semi--empirical SEDs of Dale \etal\
(2001).  It is important to note that there is some scatter between
far--IR colors and total IR luminosity, which is not included in the
figure (see Chapman \etal\ 2003).  Much of this scatter can be reduced
by including \spitzer\ 70~\micron\ data to constrain the
mid--to--far-IR `color' (see, \eg Papovich \& Bell 2002).

\textit{IR--luminous galaxies are readily 
identified out to $z\sim 1.3$.} Galaxies at higher redshifts generally
lie beyond the COMBO-17 limits.   At low redshifts ($z\lsim 0.2$) most
of the 24~\micron--selected sources correspond to relatively normal
star--forming galaxies with some starbursts.  This reflects the
limited volume probed by the survey for these redshifts ($\sim
14000$~Mpc$^3$), in which few IR--luminous galaxies are expected.  The
majority of 24~\micron\ sources with $z\sim 0.4-1$ correspond to
LIRGs, and these sources likely dominate the IR luminosity density at
these redshifts.  ULIRGs are generally not common in this field until
$z\gsim 0.8$, and LIRGs generally seem to dominate the IR emission at
these redshifts as well.   ULIRGs appear scarce even at these high
redshifts, although we are certainly missing optically faint ULIRGs
with $R \gsim  24$ (see, \eg Egami \etal\ 2004).  To study their
properties will require both large survey areas and deep optical data.

\begin{figure}
\centerline{\psfig {figure=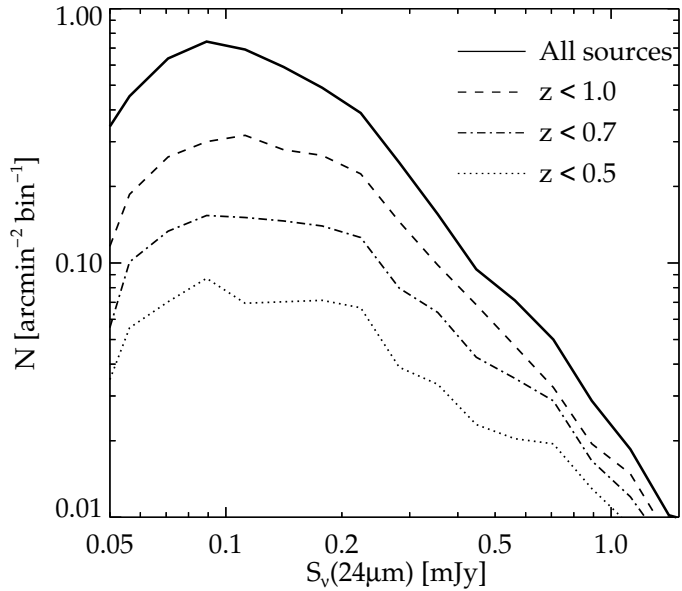,height=3.7in}}
\caption{Differential \spitzer\ 24~\micron\ number counts in the
  region of the CDF-S field covered by COMBO--17, which covers
  roughly  900~sq.\ arcmin.   The solid curve shows the counts from
  all 24~\micron\ sources in this area.  The broken lines show the
  contribution to the counts from optically--selected galaxies with
  redshifts below 0.5 (dotted line), 0.7 (dot--dashed line), and 1.0
  (dashed line).  Each bin is $\Delta (\log S_\nu) =
  0.1$~dex. Optically selected IR galaxies with redshifts $z < 1.0$
  contribute roughly half of the 24~\micron\ source counts at the
  faint end ($0.1-0.4$~mJy).   \label{fig:redshiftcounts}}
\end{figure}

Figure~\ref{fig:redshiftcounts} shows the 24~\micron\ differential
number counts in the \spitzer--COMBO-17 overlap areas, and the
contribution of galaxies as a function of redshift.  At the bright
end, most of the counts are due to galaxies with $z\lsim 1$.   At
fainter 24~\micron\ flux densities, higher--redshift galaxies dominate
the counts. Galaxies with $z\lsim 0.7$ make up only one--quarter to
one--third of the total counts at $\sim 0.2$~mJy (near the peak in
figure~\ref{fig:counts}).   Similarly, 24~\micron\ sources at these
flux densities with $z\lsim 1$ contribute only $\sim 50$\% of the
total counts.    There is roughly one \spitzer\ 24~\micron\ source per
sq.\ arcmin with no counterpart in the optical images to $R\sim 25$.
Elbaz \etal\ (2002) found a redshift distribution of $\textit{ISO}$
15~\micron\ sources with a median at $z\sim 0.7$ and a small tail to
$z\sim 1$.  In contrast, the \spitzer\ 24~\micron\ data is very
senstive to galaxies at $z\gsim 1$.  \textit{The \textit{ISO}
populations make up only a fraction of the faint 24~\micron\ sources.}

\subsection{Evolution of the IR--luminosity density}

The available redshifts allow a crude estimate for the evolution in
the IR luminosity density relative to that in rest--frame UV and
visible bands in the CDF--S.  From $z\sim 0.2-1$, the luminosity
density in the rest--frame $U$ and $V$ bands increases by roughly a
factor $\sim 3$ (uncorrected for extinction effects or incompleteness
effects), consistent with findings from previous studies (\eg Lilly
\etal\ 1996).  In comparison, the IR luminosity density grows by
roughly a factor $\gsim 8$, where the inequality symbol denotes the
fact that this estimate does not include the
contribution from IR--luminous galaxies fainter than the magnitude
limits ($R\sim 24$) of the ground--based surveys.  This underlines the
fact that \textit{the IR--luminous galaxy population appears to evolve more
rapidly than that directly measured from UV/optical--selected galaxies. }

\section{\hst\ Observations of Distant \spitzer\ Galaxies}

\begin{figure}
\centerline{\psfig {figure=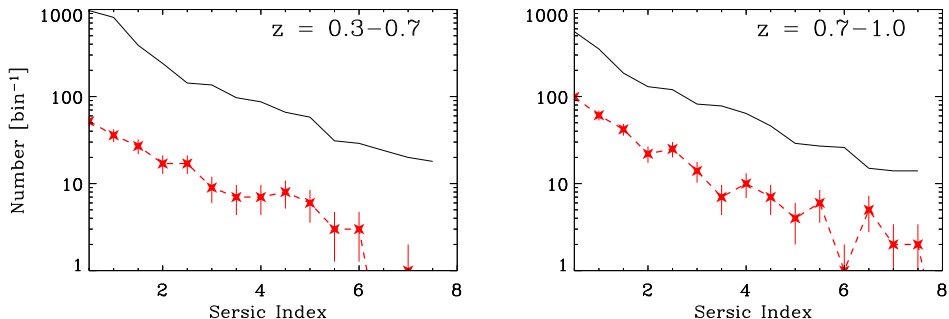,width=5.3in}}
\caption{Distribution of Sersic indices for GEMS--selected galaxies
  based on two--dimensional fits to the F850LP images (see Peng
  \etal\ 2002).  The panels show the distribution for   galaxies with
  redshifts between $z=0.3-0.7$ and $z=0.7-1.0$ (as labeled).  The
  stars connected by dashed lines show the distribution of Sersic
  indices for IR--luminous galaxies with $L_\mathrm{IR} >
  10^{10.5}$~\lsol.  The IR--luminous galaxies show approximately the
  same distribution of Sersic indices as the general galaxy population
  in both redshift intervals, which suggests that these morphological
  parameters are not indicative of IR--active stages of galaxy
  evolution.\label{fig:sersic}}
\end{figure}

The Advanced Camera for Surveys (ACS) has greatly improved the
efficiency of imaging with \hst. The region of the CDF--S has
extensive \hst/ACS imaging from the Galaxies Evolution through 
Morphologies and SEDs survey (GEMS; Rix \etal\ 2004), which provides
F606W and F850LP imaging over roughly 780 sq.\ arcmin, i.e., most of
the COMBO-17 field.\footnote{The Great Observatories Origins Deep
Survey has deeper ACS imaging in a smaller $160$~sq.\ arcmin area
within the GEMS field; see Giavalisco \etal\ (2004).}  To date, \hst\
is the most efficient means of obtaining kpc--scale resolution of
distant galaxies, as terrestrial adaptive--optics techniques are
currently only effective over small patches of sky.  Combined with the
high--quality redshifts, these wide--area ACS data allow us to test
whether structural properties and environmental effects correlate with
IR--luminous stages of galaxy evolution. 

\textit{What is the distribution of galaxy morphological
types that are in IR--active evolutionary stages?}  As a first
experiment, one can parameterize morphological type from
the \hst\ images simply in terms of
the Sersic index, $n_s$ (also called the generalized
de~Vaucouleur profile), where $I(R) \sim \exp( -R^{1/{n_s}} )$.
Objects with exponential surface brightness profiles have Sersic
indices $n_s \sim 1$, which is typical of disk--like galaxies.
Objects with more concentrated surface--brightness profiles have
higher Sersic indices, as in the case of
spheroids and bulges.  Classical $r^{1/4}$--law galaxies have $n_s = 4$. 
Crudely speaking the Sersic index quantifies the bulge--to--disk
ratio of a galaxy's light emission, and it can be used to discriminate
between late--type, disk--dominated galaxies ($n_s \leq 2.5$), and
early--type, bulge--dominated galaxies ($n_s > 2.5$).  

Figure~\ref{fig:sersic} shows the distribution of Sersic indices for
all galaxies in the GEMS catalogs with redshifts $z = 0.3-0.7$ and
$0.7-1.0$.  The distribution is skewed towards large numbers of
galaxies with lower Sersic indices, which illustrates the fact that
the majority of galaxies are disk--dominated.
\textit{Interestingly, the distribution of Sersic indices for the IR--luminous
galaxies, $L_\mathrm{IR} \geq 10^{10.5}$~\lsol, is nearly identical to that of
the general galaxy population regardless of redshift.}  The implication
is that these morphological indicators alone are a poor
discriminator of IR--activity.
%Another observation is that while
%the distribution is roughly the same, a higher fraction of sources in
%the $z\sim 0.7-1.0$ are in an IR--luminous stage of evolution relative
%to galaxies at $z\sim 0.3-0.7$.   

\begin{figure}
\centerline{\psfig {figure=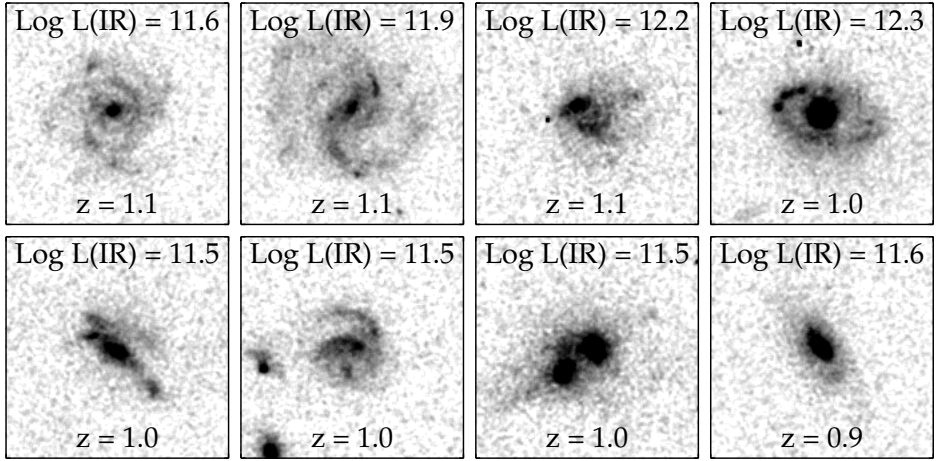,width=5.5in}}
\caption{ACS F850LP images of luminous IR galaxies at $z \sim 1$ in
  the CDF--S/GEMS areas.  Each image is 5~arcsec per side, which
  is roughly 40~kpc at these redshifts.  Total IR
  luminosities are given in units of log \lsol.
  The galaxies have a wide range of morphologies.  Clear
  distortions are evident (the two bottom, leftmost galaxies),
  including rings (upper rightmost galaxy), and multiple nuclei
  (second from right, bottom row).  However, the two
  leftmost galaxies on the top row appear to be fairly regular
  late--type spirals, and the right-most galaxy on the bottom row
  appears to have a normal spheroidal component.
  \label{fig:images}} 
\end{figure}

Even the most luminous IR galaxies, $L_\mathrm{IR} \gsim
10^{11.5}$~\lsol, span a range of morphological type.  As an illustration,
figure~\ref{fig:images} shows the ACS/F850LP images of several
fiducial IR--luminous galaxies at $z\sim 1$ from the GEMS data.  At
these redshifts the F850LP filter probes roughly the rest--frame
$B$-band of these galaxies.  It is clear that while many of these
types of galaxies have highly disturbed morphologies or evidence of
strong mergers, there are also clear examples of fairly normal galaxy
types.  Systematic studies using these data at $z\sim 0.7$
(E.\ Bell \etal\, in preparation), and as a
function of IR--luminosity (C.\ Papovich \etal, in preparation) will help
to understand the relation between galaxy morphology, environment, and
IR--activity. 

\section{Selection of Deep, Extragalactic Survey Fields}

We close with a discussion on how to choose the
location of deep fields for studying IR--luminous galaxies.  The
dominant sources of IR background are zodiacal light
and emission from cirrus clouds in the Milky Way.  Zodiacal light
dominates at mid--IR wavelengths ($3-40$~\micron),
decreases rapidly with ecliptic latitude, and shows strong seasonal
changes (\eg Price \etal\ 2003). Galactic cirrus is the dominant source of
background at far--IR wavelengths ($40-200$~\micron).  It scales
roughly linearly with the Galactic column density,
$N(\mathrm{HI})$ (Lockman \etal\ 1986;
Boulanger \& Perault 1988), and produces two noise components for IR
observations.  The first is due simply to the elevated sky brightness,
which limits the flux sensitivity of an observation by a factor roughly
$\sim \sigma_\mathrm{cirrus}^{-1}$. Fields near the plane will
require exposures several times longer than fields near the
poles to achieve comparable depth.  
%This is similar to limitations on
%X-ray observations, which are also sensitive to the total $N(\mathrm{HI})$.  

The second component is confusion noise from structure within cirrus
clouds (\eg Low \etal\ 1984; Helou \& Beichman 1990; Kiss \etal\ 2001,
2003).  Helou \& Beichman expressed the cirrus confusion noise as
$\sigma_{\mathrm{cirrus}} \sim \lambda^{2.5} D^{-2.5}
B_\lambda^{1.5}$, where $\lambda$ is the emitted wavelength, $D$ is
the diameter of the telescope aperture, and $B_\lambda$ is the mean
sky brightness. Because cirrus brightness correlates with
Galactic hydrogen column density, an increase in $N(\mathrm{HI})$ by a
factor of five corresponds to an increase in the relative confusion
noise by a factor of ten. Far--IR observations in fields with high
cirrus sky brightness pay a substantial penalty in terms of
cirrus--confusion noise, and this imposes a hard limit on the
ultimate survey depth in such fields.

\begin{figure}
\centerline{\psfig {figure=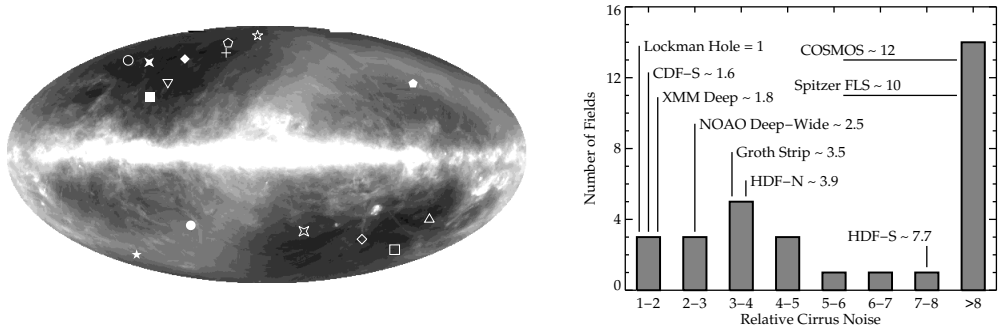,width=5.5in}}
\caption{\textit{Left}: Location of selected extragalactic survey
  fields on the Galactic \textit{IRAS} 100~\micron\ image.  Galactic north is
  up, and east to the left.  Galactic cirrus dominates the image,
  although zodiacal emission is evident and runs roughly from the
  bottom left to the upper right.  The symbols denote fiducial fields:
  HDF--N, \textit{filled star}; HDF--S, \textit{open star}; CDF--S,
  \textit{open square}; Marano, \textit{open diamond}; XMM-Deep,
  \textit{open pentagon}; XMM-LSS, \textit{filled pentagram}; NOAO
  Bo\"otes, \textit{cross}; Groth strip, \textit{filled  diamond};
  ELAIS-N1, \textit{downward triangle}; ELAIS-S2, \textit{upward
  triangle}; COSMOS, \textit{pentagon}; Lockman Hole, \textit{open
  circle}; SSA22, \textit{filled circle}; Subaru Deep, \textit{open
  pentagram}; \spitzer\ FLS, \textit{filled square}. \textit{Right}:
  Far--IR cirrus noise of the extragalactic survey fields listed in
  table~1 relative to that of the Lockman Hole.  Estimates are based
  on the Galactic $N(\mathrm{HI})$, which correlates linearly with
  cirrus brightness.  High cirrus noise substantially increases the
  confusion noise at all angular scales and is a strongly limiting
  factor for far--IR observations. \label{fig:iras100um}}
\end{figure}

Table~1 lists the properties of known extragalactic survey fields
(updated and adapted from a compilation by Stiavelli \etal\ 2003).
For each field, the Galactic extinction (parameterized by the color
excess, $E(B-V)$) and hydrogen column density are taken from the maps
of Schlegel \etal\ (1998) and Dickey \& Lockman (1990), respectively.
The left panel of figure~\ref{fig:iras100um} shows the location of
several of these fiducial fields superimposed on an \textit{IRAS}
100~\micron\ all-sky image.  The right panel of
figure~\ref{fig:iras100um} shows the distribution of Galactic cirrus
confusion noise relative to that of the Lockman Hole, the sightline
with the minimum $N(\mathrm{HI})$.

\textit{Fields that are both far from the ecliptic (low zodiacal light) and
the Galactic plane (low $N(\mathrm{HI})$ and cirrus) have the
lowest backgrounds and confusion noise, and are, in a sense, chosen by
nature to be the ideal locations for full multi--wavelength
extragalactic surveys.}  Future IR missions (\eg JWST, Herschel, SAFIR)
will gravitate to these fields, as well as future X--ray telescopes (\eg
Constellation--X).  To optimally study IR--luminous stages of
galaxy evolution will require full panchromatic surveys in these fields,
including high--angular \hst\ imaging.  These multi--wavelength data
will be crucial for dissecting the mechanisms for galaxy evolution,
not only in the \spitzer\ era, but for decades to come. 

\tabcolsep 0.1in 
\begin{table}
\center{Compilation of the Properties of Known Extragalactic Survey Fields.}
\begin{tabular}{lcccccc}
 & R.A. & Decl. & $l$ & $b$ & &
$N(\mathrm{HI})$ \\ 
\multicolumn{1}{c}{Name} & (J2000.0) & (J2000.0) & (deg) & (deg) & $E(B$$-$$V)$ &
($10^{20}$cm$^{-2}$) \\ 
\hline
DEEP-1     & \phn 0:17:00 & \phs16:00:00.0 &  111.0 &  -46.1 & 0.049 & 4.18 \\
WHT Deep   & \phn 0:22:33 & \phs 00:20:57.0 &  107.6 &  -61.7 & 0.025 & 2.73 \\
ELAIS-S1   & \phn 0:34:44 & -43:28:12.0 &  313.5 &  -73.3 & 0.008 & 2.52 \\
FORS Deep  & \phn 1:06:04 & -25:45:46.0 &  191.1 &  -86.5 & 0.018 & 1.88 \\
XMM-LSS    & \phn 2:21:20 & -04:30:00.0 &  170.3 &  -58.8 & 0.027 & 2.61 \\
CNOC2      & \phn 2:23:00 & \phs 00:00:00.0 &  165.7 &  -55.1 & 0.039 & 2.96 \\
DEEP-2     & \phn 2:30:00 & \phs 00:00:00.0 &  168.1 &  -54.0 & 0.022 & 2.89 \\
CFDF       & \phn 3:00:00 & \phs 00:00:00.0 &  177.0 &  -48.9 & 0.096 & 6.99 \\
Marano     & \phn 3:15:09 & -55:13:57.0 &  270.2 &  -51.8 & 0.016 & 2.45 \\
CDF-S$^\dag$ & \phn 3:32:30 & -27:48:47.0 &  223.6 &  -54.4 & 0.008 & 0.79 \\
ELAIS-S2   & \phn 5:02:24 & -30:35:55.0 &  232.6 &  -35.7 & 0.012 & 1.43 \\
CNOC2      & \phn 9:20:00 & \phs 37:00:00.0 &  186.6 & \phs  44.7 & 0.011 & 1.47 \\
COSMOS     & 10:00:29 & \phs 02:12:21.0 &  236.8 &  \phs 42.1 & 0.017 & 2.90 \\
Lockman Hole$^\dag$ & 10:52:43 &  \phs57:28:48.0 & 149.3 & \phs 53.1 & 0.008 & 0.57 \\
EIS Deep   & 11:20:45 & -21:42:00.0 &  276.4 &  \phs 36.5 & 0.045 & 4.18 \\
HDF-N$^\dag$      & 12:36:49 &  \phs62:12:58.0 &  125.9 & \phs  54.8 & 0.012 & 1.41 \\
SSA13      & 13:12:21 &  \phs42:41:21.0 &  109.0 &  \phs 73.9 & 0.014 & 1.46 \\
Subaru Deep & 13:24:21 &  \phs27:29:23.0 &  \phn37.6 & \phs  82.7 & 0.019 & 1.19 \\
XMM Deep$^\dag$   & 13:34:37 & \phs37:54:44.0 & \phn85.6 & \phs75.9 & 0.006 &
0.83 \\
Groth Strip$^\dag$ & 14:16:00 &  \phs52:10:00.0 &  \phn 96.3 & \phs  60.4 & 0.013 & 1.30 \\
ELAIS-N3   & 14:29:06 &  \phs33:06:00.0 &   \phn54.7 &  \phs 68.1 & 0.008 & 1.11 \\
NOAO Bo\"otes$^\dag$ & 14:32:06 &  \phs34:16:47.5 & \phn  58.2 &  \phs 67.7 & 0.012 & 1.04 \\
ELAIS-N1   & 16:10:01 &  \phs54:30:36.0 &   \phn84.3 &  \phs 44.9 & 0.005 & 1.38 \\
ELAIS-N2   & 16:36:58 &  \phs41:15:43.0 &   \phn65.3 &  \phs 42.2 & 0.007 & 1.07 \\
DEEP-2       & 16:52:00 &  \phs34:55:00.0 & \phn  57.4 &  \phs 38.3 & 0.016 & 1.78 \\
\spitzer\ FLS & 17:18:00 &  \phs59:30:00.0 &  \phn 88.3 &   \phs34.9 & 0.023 & 2.66 \\
CFHT Legacy & 22:15:31 & -17:44:05.0 &  \phn39.3 &  -52.9 & 0.026 & 2.39 \\
SSA22      & 22:17:35 &  \phs00:15:30.0 &   \phn63.1 &  -44.0 & 0.066 & 4.64 \\
DEEP-2     & 23:30:00 &  \phs00:00:00.0 &   \phn85.0 &  -56.7 & 0.037 & 4.04 \\
HDF-S      & 22:32:56 & -60:30:02.7 &  328.3 &  -49.2 & 0.027 & 2.22 \\
EIS Deep   & 22:50:00 & -40:12:59.0 &  357.5 &  -61.7 & 0.011 & 1.47 \\
\hline
\end{tabular}
$^\dag$~Denotes field included in the \spitzer\ GTO cosmological surveys.
\end{table}

\begin{acknowledgments}
We are indebted to all our collaborators on this project for their
work and for allowing us to present some of this material prior to
publication.  We wish to acknowledge the fellow members of the
\textit{Spitzer} MIPS, IRAC, and IRS GTO teams for stimulating
conversations, efficient processing and analysis of the data, and much
hard work.  We would also like to thank the members of the COMBO--17
and GEMS teams for their assistance and useful conversations, in
particular Eric Bell, Dan McIntosh, Hans--Walter Rix, Rachel
Somerville, and Christian Wolf.   C.P.\ acknowledges highly
interesting conversations with other participants at the Aspen Center
for Physics, where much of this work was completed.  Finally, we would
like to extend our deep appreciation to the symposium organizers for
the invitation to present this material, and for planning such an
interesting and successful meeting.  Support for this work was
provided by NASA through contract 960785 issued by JPL/Caltech.
\end{acknowledgments}

\end{document}